\pdfoutput=1
\documentclass[useAMS,usenatbib]{mnras}
\usepackage{graphicx}
\usepackage{amsbsy,amsmath}
\usepackage{graphics,graphicx}
\usepackage{subfloat}
\usepackage[compatibility=false]{caption}
\usepackage[skip=0.1pt]{subcaption}
\usepackage{ae,aecompl}
\usepackage[compact]{titlesec}
\usepackage{notoccite}
\usepackage{caption}
\usepackage{subcaption}
\usepackage{amsbsy,amsmath}
\usepackage{graphics,graphicx}
\usepackage{longtable}
\usepackage[latin1]{inputenc}
\usepackage{amsmath}
\usepackage{amsfonts}
\usepackage{amssymb}
\usepackage{makeidx}
\usepackage{float}
\restylefloat{figure}
\usepackage{stfloats}
\usepackage{epsfig}
\usepackage{graphicx}
\usepackage{caption}
\usepackage{color}
\usepackage{tabularx}
\usepackage{natbib}
\usepackage{multicol}
\usepackage{gensymb}
\usepackage[toc]{appendix}
\usepackage{breqn}
\usepackage{newunicodechar} 

\title{Substructure analysis of the RXCJ0232.2-4420 galaxy cluster }

\author[Parekh et al.]{{Viral Parekh$^{1,2}$\thanks{E-mail: vparekh@ska.ac.za}, Tatiana F. Lagan\'{a}$^{3}$, and Ruta Kale$^{4}$}
\\
$^{1}$Department of Physics and Electronics, Rhodes University, Drosty Rd, Grahamstown, 6139, Republic of South Africa\\
$^{2}$South African Radio Astronomy Observatory (SARAO),  2 Fir Street, Black River Park, Observatory, 7925, South Africa \\
$^{3}$N\'ucleo de Astrof\'\i sica, Universidade Cidade de S\~ao Paulo, R. Galv\~ao Bueno 868, Liberdade, S\~ao Paulo, SP, 01506-000, Brazil\\
$^{4}$National Centre for Radio Astrophysics, Tata Institute of Fundamental Research, Savitribai Phule Pune University Campus, \\Pune 411 007 Maharashtra, INDIA\\
}

\date{MNRAS Accepted.}

\pubyear{2015}

\begin{document}
\label{firstpage}
\pagerange{\pageref{firstpage}--\pageref{lastpage}}
\maketitle

\begin{abstract}
{ RXCJ0232.2-4420, at $z$ = 0.28, is a peculiar system hosting a radio halo source around the cool-core of the cluster. To investigate its formation and nature, we used archival {\it Chandra} and XMM-\textit{Newton} X-ray data to study the dynamical state of the cluster and detect possible substructures in the hot gas. Its X-ray surface brightness distribution shows no clear disruption except an elongation in the North-East to South-West direction. We perform the unsharp masking technique and compute morphology parameters (Gini, $M_{20}$ and concentration) to characterise the degree of disturbance in the projected X-ray emission. Both of these methods revealed a substructure, which is located at $\sim$ 1$'$ from the cluster core in the South-West direction. {Previous spectral analysis conducted for RXCJ0232.2-4420 concluded that there are a short cooling time and low entropy at the cluster centre, indicating that the cluster has a cool core.} Thus, we suggest that RXCJ0232.2-4420 may be a system where the core of the cluster is not showing any sign of disturbance, but the South-West substructure could be pumping energy to the detected radio halo via turbulence.}

\end{abstract}

\begin{keywords}
Clusters of galaxies; intra-cluster medium, X-ray clusters
\end{keywords}

\section{Introduction}
\par Galaxy clusters are the most massive bound and quasi-relaxed objects in the Universe. In the mass composition of galaxy clusters, the majority of the baryonic material is in the form of hot and diffuse gas, namely the intra-cluster medium (ICM). This hot ICM retains the cluster's  {thermodynamic properties as well as imprints of recent merger} history. Galaxy clusters can be broadly divided into two categories - relaxed and non-relaxed clusters.  Generally, relaxed clusters possess a cool-core (CC) while non-relaxed clusters (NCC) have either weak cool-core or no cool-core \citep[but see][]{Lagana19}. {However, there is not an unambiguous procedure that distinguishes these two classes of clusters with a well-defined separation between NCC and CC clusters} \citep{2017ApJ...846...51L,2015A&A...575A.127P,2013A&A...549A..19W}.


The dynamic classification of a cluster has important implications both for using clusters as cosmological tools and for studying the complex gravitational and non-gravitational processes acting during large-scale structure formation and evolution. 
Furthermore, the measurements of the global properties of clusters, for example, hydrostatic mass and scaling relations also depends on the cluster dynamical state \citep{2020ApJ...892..102L,2009A&A...498..361P}.

\par {Galaxy clusters also allow us to study large-scale and non-thermal diffuse radio sources such as radio halos and relics which are believed to be linked with substructures and cluster merging process} \citep{2010ApJ...721L..82C,2009A&A...507..661B}. 
{These diffuse radio sources are generally classified based on their shapes and locations in the cluster: radio halos which are mostly circular and situated at the cluster centre, while radio relics are found at the periphery of the cluster and are elongated perpendicular to the merger axis} \citep[e.g.][]{2019SSRv..215...16V}. These diffuse radio sources trace the cluster magnetic field distribution and reservoir of relativistic electrons. The radio halos are further divided into two categories, namely giant halos (of the size of $\sim$ 1 Mpc) and mini-halos (of the size of $\sim$ 500 kpc) \citep[e.g.][]{2014IJMPD..2330007B,2012A&ARv..20...54F}. Sometimes the physical size of the radio halos is ambiguous and depends on the sensitivity and quality of radio observations. The formation mechanisms of these two classes of radio halos {are} still under debate, but from the observational point of view, they are largely connected with the dynamical state of the host cluster. Most of the giant radio halos are found in the disturbed or non-relaxed clusters, while mini-halos are associated with relaxed or cool-core clusters \citep{2010ApJ...721L..82C,2001ApJ...553L..15B}. Cluster dynamical states are playing a vital role to investigate the nature of diffuse radio sources, and recently, several puzzling large-scale size ($\sim$ Mpc) radio halos are reported in cool core clusters  \citep{2019MNRAS.486L..80K, 2016MNRAS.459.2940K, 2014MNRAS.437.2163S,2014MNRAS.444L..44B}. But still, the role of the cluster dynamics on the formation of radio halos is poorly understood.      
\par High-resolution X-ray observations provide the quantitative and robust measure of the cluster disturbance as well as the identification of subclusters, bimodality and X-ray centroid shifts in clusters map using morphology parameters \citep{2015A&A...575A.127P,2010A&A...514A..32B,2006MNRAS.373..881P,1995ApJ...439...29B,1995ApJ...452..522B}. In the literature, there are other imaging and filtering techniques applied to investigate substructures, for example, the unsharp masking technique. Some authors have also advocated studying the coincidence between the brightest cluster galaxy and X-ray peak to evaluate cluster dynamical states \citep{2018MNRAS.478.5473L, 2016MNRAS.457.4515R,2012MNRAS.420.2120M}. Cluster central cooling time and entropy are also very important parameters to investigate the cool-core clusters \citep{2010A&A...513A..37H,2009ApJS..182...12C}. Entropy governs the global property of the clusters and records the past heating and cooling history of these systems. It is very crucial to investigate excess heat generation in the ICM due to substructure or cluster merging. 
\par Some substructures, however, are less evident in the X-ray surface brightness distribution than the temperature distribution \citep{2009ApJ...696.1029A,2004ApJ...605..695G}. 
The merging process generates disturbances in the ICM, which results in the swirling of stripped gas,  shock expansion and propagation in and out of the cluster. Such shock propagation and the resulting turbulence can create a wide variety of phenomenon in the ICM. These imprints of substructures or cluster merging process are largely studied by the computation of the thermodynamic properties \citep{2016MNRAS.459.2940K,2006A&A...446..417F,2004ApJ...615..181H}. 

\subsection{RXCJ0232.2-4420}
\par RXCJ0232.2-4420 hosts a radio halo whose properties are similar to the mini-halos \citep{2019MNRAS.486L..80K}. { In the GMRT 606 MHz observation, \cite{2019MNRAS.486L..80K} estimated the total linear extent of the halo is 550 $\times$ 800 kpc$^{2}$ and the 1.4 GHz radio power of the halo is 4.5 $\times$ 10$^{24}$ W Hz$^{-1}$}.  Furthermore, they argued that the detected radio halo in RXCJ0232.2-4420 is in the transition phase of mini-halo to giant radio halo. To understand how the radio halo formed in RXCJ0232.2-4420, it is important to investigate the dynamical state of the cluster. This also requires to study of how the possible substructure has triggered the particle re-acceleration process in the ICM. 
\par The dynamical state of RXCJ0232.2-4420 is not clear in the literature. Two independent parametric substructure analysis was performed for RXCJ0232.2-4420 based on the X-ray observations: \cite{2013A&A...549A..19W} have used a sample of 129 clusters observed by XMM-\textit{Newton} to investigate the substructure occurrence as a function of redshift. They have used two shape estimators, namely power ratio ($P3/P0$) and centroid shift ($w$) to quantify the dynamical state of the clusters. Based on their analysis, they found that the RXCJ0232.2-4420 is a moderately disturbed cluster. Later on, \cite{2017ApJ...846...51L} used eight morphology parameters for the Planck Early Sunyaev-Zeldovich (ESZ) clusters that were observed by  XMM-\textit{Newton} telescope. They have shown that the identification of relaxed clusters using a set of morphology parameters is relatively easier than identifying disturbed clusters. They also show that combining the concentration parameter (which is sensitive to the core) with centroid shift (which is sensitive to substructure) gives better results to characterise the dynamical state of the system.
Based on their analysis, they found RXCJ0232.2-4420 is a relaxed system. 

\par
\begin{center}
\begin{figure*}
    \centering
    \begin{subfigure}[t]{0.45\textwidth}
        \includegraphics[width=1\textwidth]{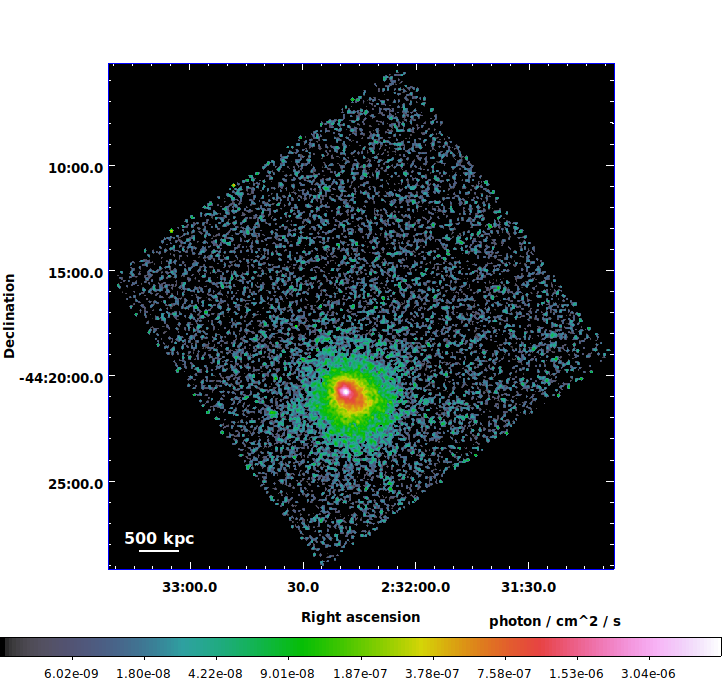}
        \caption{}
        \label{rfidtest_xaxis1}
    \end{subfigure}
    \begin{subfigure}[t]{0.45\textwidth}
        \includegraphics[width=1\textwidth]{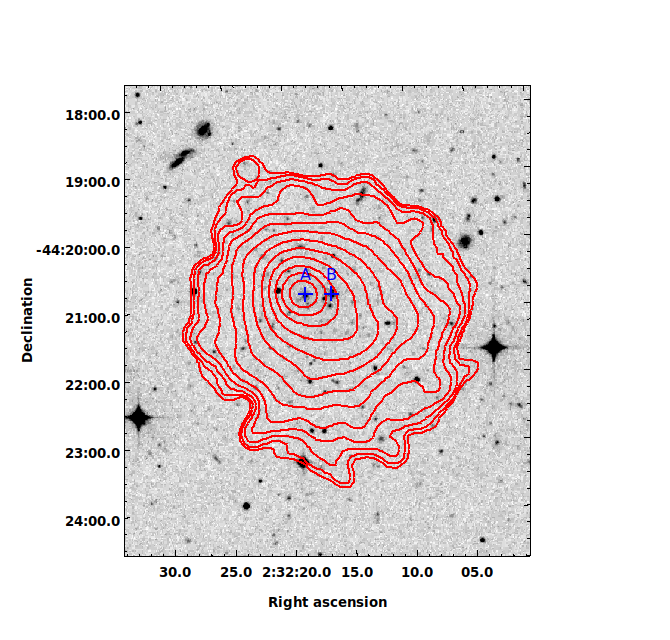}
        \caption{}
        \label{rfidtest_yaxis2}
        \end{subfigure}
    \begin{subfigure}[t]{0.45\textwidth}
        \includegraphics[width=1\textwidth]{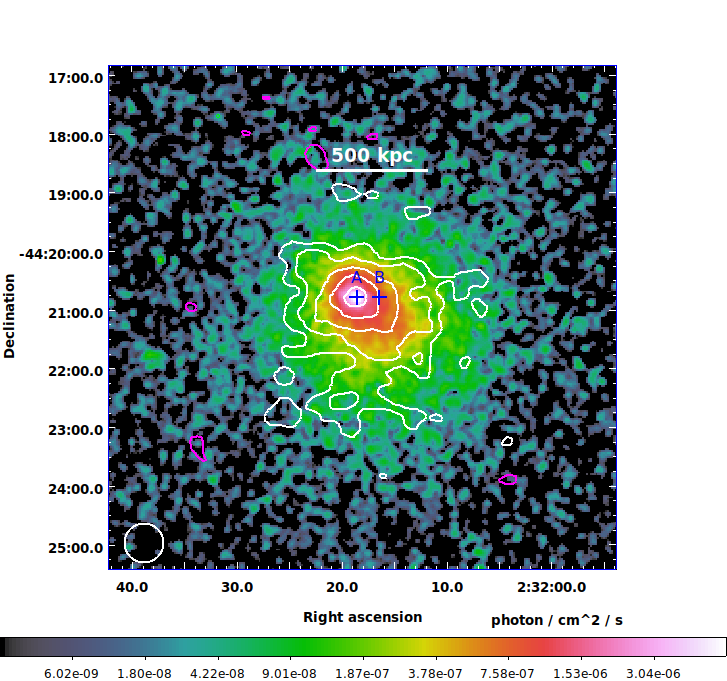}
        \caption{}
        \label{rfidtest_yaxis2}
        \end{subfigure}

    \caption{(a) Exposure corrected and background subtracted {\it Chandra} X-ray image of RXCJ0232.2-4420. (b) X-ray contours (red) overlaid on the superCOSMOS optical image. We mark the BCGs positions with the blue `+'. (c) GMRT 606 MHz radio contours overlaid on the X-ray {\it Chandra} image. The contours levels are the same as in ~\protect\cite{2019MNRAS.486L..80K}. 1$\sigma$ rms = 0.1 mJy beam$^{-1}$, resolution of 20$''$ $\times$ 20$''$. The first contour is drawn at $\pm$3$\sigma$ and increases with a factor of two. White contours are positive and magenta contours are negative. All X-ray images have energy range of 0.5-7 keV and 1.968$''$ pixel size.}
    \label{x-ray img}        
\end{figure*}
\end{center}

 We show {\it Chandra} X-ray  and optical superCOSMOS images for RXCJ0232.2-4420 in Fig.~\ref{x-ray img}(a) and (b), respectively, and we list its X-ray properties in Table \ref{sample}. 
There are two brightest cluster galaxies (BCGs) found in the core of the cluster \citep{2008A&A...483..727P}. In the optical image, we marked their positions with the blue symbol `+'. In Fig.~\ref{x-ray img}(c), diffuse radio halo contours are shown to be surrounding one of the BCGs at the cluster core, which is a typical signature of a mini-radio halo.  
 Based on the visual inspection of the {\it Chandra} X-ray image, it looks like a regular and relaxed cluster with symmetric contours around the X-ray peak. There are no merging activities visible as compared to other disturbed clusters hosting radio halos. However, it shows the broad extension in the South-West direction. One of the BCGs is spatially coincident with the X-ray peak as for a typical relaxed system.

\par In this paper, we have quantitatively investigated the dynamical state of this cluster. This will help us to identify any underlying substructure and its connection with the detected radio halo. This paper is organised as follows; Section~2 gives X-ray data analysis, Section~3 presents the results, Sections~4 and 5 are the discussion and conclusion, respectively. Throughout this paper, we assume $H_{0}$ = 70 km s$^{-1}$ Mpc$^{-1}$, $\Omega_{M}$ = 0.3 and $\Omega_{\Lambda}$ = 0.7. At redshift $z$ = 0.2836, 1$''$ corresponds to 4.28 kpc and a luminosity distance = 1456 Mpc.  In this paper, all confidence intervals correspond to 1$\sigma$.

\begin{table*}
\centering
\begin{small}
\caption{RXCJ0232.2-4420 X-ray properties from the MCXC catalogue \citep{2011A&A...534A.109P}.}
\begin{tabular}{ccccccccccccccc}
\hline
\hline
Cluster Name & RA(J2000) & DEC(J2000) & $z$ & $L_{\rm X}[0.1-2.4] \rm \,keV$ & $M_{500}$   \\
    & h~~m~~s  & h~~m~~s & & ($10^{44} \rm \, erg/s$) & (10$^{14}$ $M_{\odot}$)  \\
\hline
RXCJ0232.2-4420  &  02~~32~~16.8 &    -44~~20~~51.0  &0.2836 &8.49     &6.13   \\

\hline
\end{tabular}
\label{sample}
\end{small}
\end{table*}

\section{X-ray observation and data analysis}

\subsection{Chandra data analysis}
\par We use {\it Chandra} archival data (obsid 4993, date of observation 8th June 2004, and total exposure time $\sim$ 23.4 ks) to study RXCJ0232.2-4420 for both imaging and spectral analysis purposes. We followed the standard X-ray data analysis steps in the data reduction procedure. We processed the {\it Chandra} data with the {\bf C}handra {\bf I}nteractive {\bf A}nalysis of {\bf O}bservations (CIAO v 4.11) \citep{2006SPIE.6270E..1VF} and CALDB 4.8.3 software. In the data reduction steps, in the beginning, we use the \verb"chandra_repro" task to reprocess the Advanced CCD Imaging Spectrometer (ACIS) imaging data, followed by removing high background flares (3$\sigma$ clipping) with the task \verb"lc_sigma_clip". All filtered event files ({exposure time $\sim$ 22,414 sec} after flare removal) are in the energy band of [0.5--7.0]\,keV in further analysis. We binned the event files with the factor of 4, which results in 1.968$''$ image pixel size. We detect and subtract point sources around the cluster emission using the CIAO's \verb"wavdetect" and \verb"dmfilth" tasks, respectively. Finally, we divided the count image by the exposure map and generated the flux image (photon~cm$^{-2}$~s$^{-1}$). We make the exposure map using the \verb"fluximage" task of the CIAO software. We also subtract the background from the flux image using the {\it Chandra} blank sky files\footnote{http://cxc.harvard.edu/ciao/threads/acisbackground/}. 

\subsection{Chandra spectral analysis} \label{chandra_spectral}
We extract the X-ray spectra using the \verb"specextract" task of the CIAO. Then we use the package Sherpa (version 1) in order to perform the {\it Chandra} spectral fitting. {In the spectral analysis procedure, spectra are grouped such that each bin has a minimum count of 10}. In all {\it Chandra} spectral fit procedure, the redshift of RXCJ0232.2-4420 is fixed at $z =$ 0.2836, metallicity (Z) at 0.3 of the solar value, and the hydrogen column density value is fixed at $n_{H}$ = 2.48 $\times$ 10$^{20}$ atoms cm$^{-2}$ estimated using the colden toolkit\footnote{https://cxc.harvard.edu/toolkit/colden.jsp} (NRAO dataset), while the temperature ($kT$) and the spectrum normalisation (K) were free to vary. A single temperature fit is used to model the cluster X-ray spectrum with an absorbed thermal plasma emission model WABS(APEC). Abundances were estimated assuming the ratios from \citet{Asplund09}. We employed {\it Chandra} blank sky event files, for background subtraction, after taking into account the background scaling method.

\subsection{XMM-\textit{Newton} Data Reduction}
\label{sec:XMM}

\par We use XMM-\textit{Newton} archival data of RXCJ0232.2-4420. The cluster was observed on 2002 July 11th for $\sim$ 13,645 sec for MOS (XMM-\textit{Newton}, ObsID:0042340301). In this observation, the prime full-frame mode was employed for the XMM-\textit{Newton}'s three cameras (extended mode for pn) with thin filters. The data reduction was done with the {\bf S}cience {\bf A}nalysis {\bf S}ystem (SAS) \citep{2004ASPC..314..759G} version 16.1.0 (July 2017) and calibration files updated to April 2019.  For the purpose of filtering background flares, we applied a 2$\sigma$-clipping procedure using the {light curves in the [10--14]\,keV energy band}. The resulting ``cleaned'' exposure times are 11.3 ks for MOS cameras and 6.4 ks for pn, respectively. Then, to take into account the background contribution for each detector, we obtained a background spectrum in an outer annulus of the observation in the [10-12] keV energy band. We matched these background spectra with the blank sky obtained by \citet{ReadPonman03} in the same energy band and region. Finally, we rescaled the observation background to the blank sky background to get a normalisation parameter that will be applied in the spectral fits. Discrete X-ray point sources were detected by visual inspection, confirmed in the High Energy Catalogue 2XMMi Source, and excluded from events files. 

\subsection{XMM-\textit{Newton} spectral analysis}
In our work, the XMM-\textit{Newton} spectral analysis was restricted to the energy range  [0.7--7.0]\,keV. Furthermore, to avoid any influence from Al and Si instrumental lines, we also exclude the energy band from 1.2 to 1.9 keV. 
All the extracted spectra are re-binned to ensure at least 15 counts in each energy bin. To model the spectrum, we adopt an absorbed thermal plasma emission model WABS(APEC) as similarly used for the {\it Chandra} spectral analysis. Abundances were also estimated considering the ratios from \citet{Asplund09}. In all the fitting procedures performed using the XSPEC version 12.9.1, the redshift and $n_{H}$ were fixed at $z =$ 0.2836 and 2.48 $\times$ 10$^{20}$ atoms cm$^{-2}$, respectively. The $kT$, Z, K were free to vary. 

If we divide the normalization by the number of pixels in each bin (K/$N_{\rm P}$), we obtain the projected emission measure (EM). EM is proportional to the square of the electron density $n_{e}^{2}$, integrated along the line of sight. Thus, with the best-fit $kT$ and K, we estimate the pseudo-entropy ($S$) and pseudo-pressure ($P$) in the following ways: $S= kT \times [\frac{K}{N_{\rm P}}]^{-1/3}$ and $P = kT \times [\frac{K}{N_{\rm P}}]^{1/2}$.

\section{Results}
\subsection{X-ray surface brightness}
\label{sb}
\par 
In this section, we derive the surface brightness profile of RXCJ0232.2-4420. We adopt the $\beta$-model \citep{1976A&A....49..137C} to study the 1D surface brightness profile.  This model is described by four parameters as shown by Eq. \ref{beta_eqn}, where $r_{c}$ is the cluster core radius (in unit of arcmin), $\beta$ controls the slope of the profile or power law index, $r$ is the reference point of the profile, and $\eta_{\rm gas}$(0) is the normalisation or the maximum value of the model, such that
\begin{equation}
\eta_{\rm gas}(r)=\eta_{\rm gas}(0)\left[1+\left(\frac{r}{r_{c}}\right)^{2}\right]^{-3\beta + 1/2} + Const.
\label{beta_eqn}
\end{equation}

\begin{center}
\begin{figure*}
    \centering
    \begin{subfigure}[t]{0.45\textwidth}
        \includegraphics[width=1\textwidth]{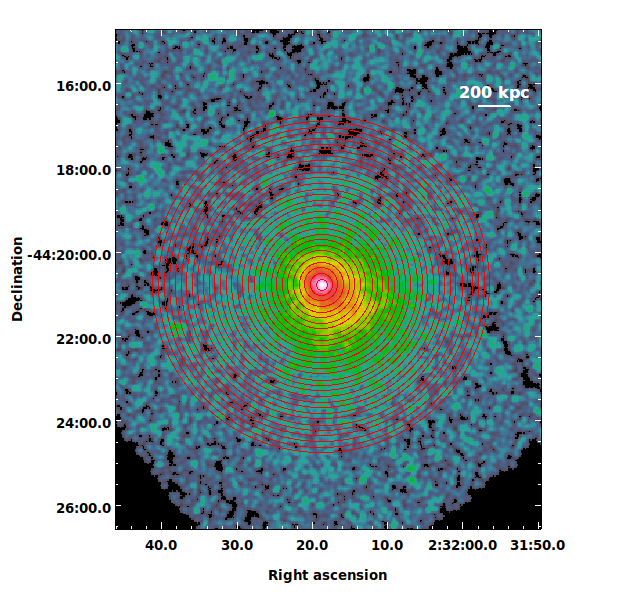}
        \caption{}
        \label{rfidtest_xaxis1}
    \end{subfigure}
    \begin{subfigure}[t]{0.50\textwidth}
        \includegraphics[width=1.1\textwidth]{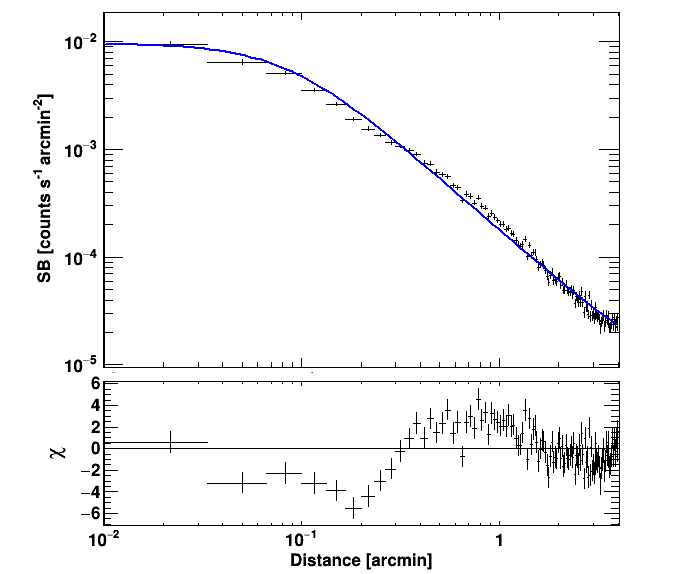}
        \caption{}
        \label{rfidtest_yaxis2}
        \end{subfigure}
    \caption{(a) Annuli centred on the X-ray peak of the cluster RXCJ0232.2-4420, used to obtain the azimuthal surface brightness profile, shown in panel (b) fitted with the $\beta$-model.}
    \label{SB_prof}        
\end{figure*}
\end{center}

We derive the azimuthal average surface brightness profile over a radius of 4$'$ ($\sim$ 1 Mpc) centred on the peak of the X-ray emission (reference point) as shown in Fig.~\ref{SB_prof}(a). We fit the surface brightness profile using the PROFFIT v1.5 software \citep{2011A&A...526A..79E}, shown in  Fig.~\ref{SB_prof}(b), assuming for each bin an SNR of at least 20.
Our best fit values and statistics are listed in Table \ref{beta_val}. At the distance of $\sim$ 1$'$ from the origin,
the best fitting profile deviates from the data points, corresponding to a large fluctuation in $\chi^{2}$, as can be seen in Fig.~\ref{SB_prof}(b).

\par
\begin{center}
\begin{figure*}
    \centering
    \begin{subfigure}[t]{0.45\textwidth}
        \includegraphics[width=1\textwidth]{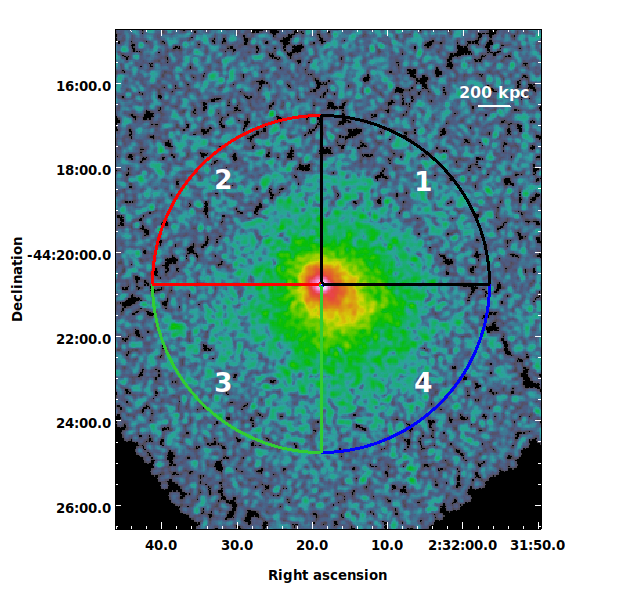}
        \caption{}
        \label{rfidtest_xaxis1}
    \end{subfigure}
    \begin{subfigure}[t]{0.50\textwidth}
        \includegraphics[width=1\textwidth]{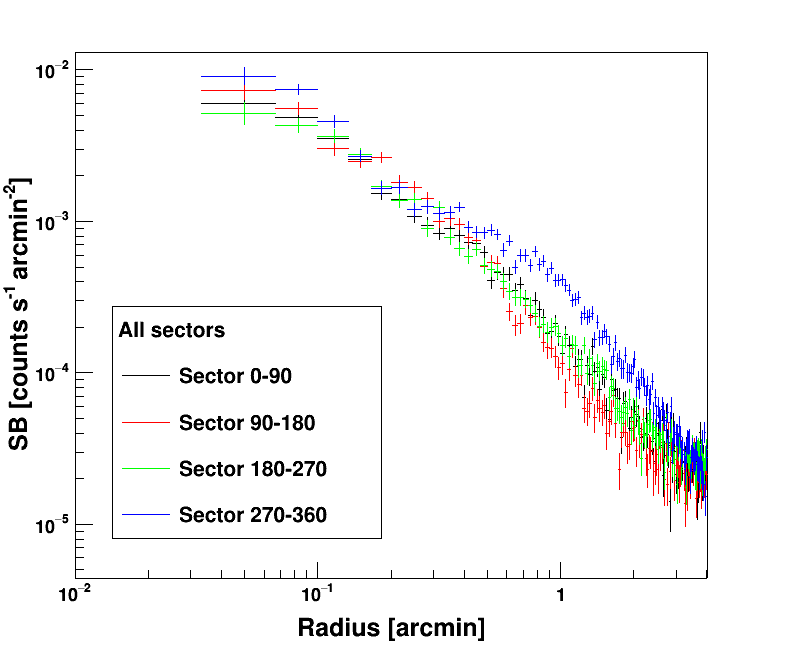}
        \caption{}
        \label{rfidtest_yaxis2}
        \end{subfigure}
    \caption{(a) Four sectors of 90$^{\circ}$ angle each centered on the X-ray peak. (b) Surface brightness profiles extracted from the four sectors, that are plotted with different colors.}
    \label{SB_prof_sec}        
\end{figure*}
\end{center}

 To investigate this deviation in the surface brightness profile, we divided the cluster region into the four 90$^{\circ}$ sectors, as shown in Fig.~\ref{SB_prof_sec}(a). For each of these sectors, we extracted the surface brightness profiles and plotted them in Fig.~\ref{SB_prof_sec}(b). As one can see, there is a visible discrepancy in sector number 4 (270$^{\circ}$ - 360$^{\circ}$ angle), in the South-West direction of the cluster, compared to the other three sectors.  

\begin{table}
\centering
\caption{ 1D $\beta$ profile and 2D $\beta$-model (see \$\ref{unshp}) best fitting values with statistics.}
\label{beta_val}
\begin{tabular}{cc}
\hline
\hline
parameter & value\\
\hline
$\beta$ & 0.43 $\pm$ 0.0014 \\
$r_{c}$ & 0.08 (22 $h_{70}$ kpc)$\pm$ 0.0012 (3 $h_{70}$ kpc)\\
Norm & 9.6 $\times$ 10$^{-3}$ \\
Const & 4.4$\times$ 10$^{-6}$ \\
\hline
$\chi^{2}$ & 412 \\
d.o.f & 118 \\
\hline 
\hline
$r_{0}$ &  60$^{+5}_{-5}$ $h_{70}$ kpc            \\
$\epsilon$ & 0.23$^{+0.012}_{-0.012}$          \\
$\theta$ & 320$^{+1.664}_{-1.664}$ deg            \\
$\alpha$ &  0.60$^{+0.014}_{-0.015}$ \\
\hline
$\chi^{2}$ & 6056\\
d.o.f & 7848 \\
\hline
\end{tabular} 
\end{table}

\subsection{Cluster residual images}
\label{unshp}
\par In order to generate the cluster's residual images, we use two different techniques: (1) unsharp masking and (2) subtract a 2D beta-model from the cluster image. In this analysis, we find that both these techniques are useful; however, the former is important to pinpoint the underlying substructure in the inner parts of the cluster, while latter is better used for revealing larger-scale features.  

\par
\begin{figure*}
\centering
\includegraphics[width=1\textwidth]{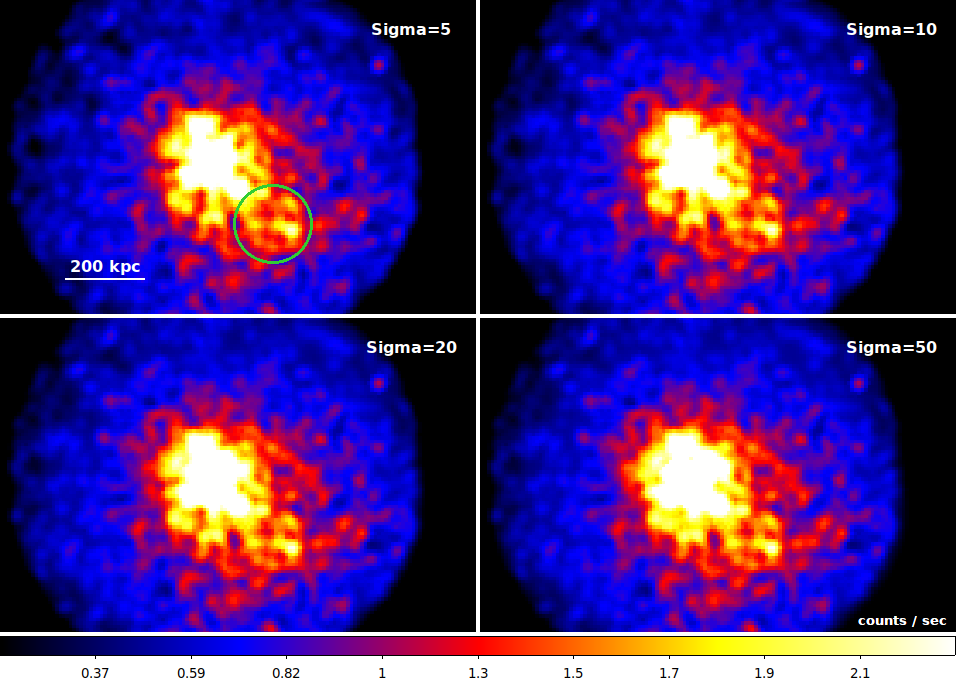}
\caption{Residual images of the unsharp mask filtering. In each image we have given the width of the Gaussian sigma used for filtering. We also located the emission excess region inside a green circle in the top-left image.}
\label{unsharp}
\end{figure*}

 Unsharp mask is a filtering technique that can be used to identify substructure within a cluster's image. This is done by smoothing a cluster emission with a filter of width $\sigma$, so that high-frequency structure can be washed out from the smoothed image. This map has only high-frequency components of the galaxy cluster's flux distribution, representing the underlying substructures. To make smooth images, we use four different Gaussian kernels of 5$\sigma$, 10$\sigma$, 20$\sigma$ and 50$\sigma$. Then, we subtract from original cluster images each of these smoothed images (and take their absolute values), and the residual images are shown in Fig.~\ref{unsharp}.  There is an excess of X-ray emission visible in the South-West direction, which is identified by the green circle as plotted in Fig.~\ref{unsharp}(a). The distance from the cluster's X-ray peak to this region of emission excess is $\sim$ 0.6$'$, and its radius is estimated to be $\sim$ 0.35$'$.  

\par We also use the 2D spherical $\beta$-model to subtract the surface brightness model from the cluster X-ray image in the [0.5--7 keV] energy band. To take into account the ellipticity of the plasma emission we use the following standard coordinates transformation: 

\begin{equation}
\begin{cases}
x^{\prime} =(x - x_{0}) \cos \theta  -  (y- y_{0}) \sin\theta \\
y^{\prime}  =(x - x_{0}) \sin \theta  -  (y- y_{0}) \cos \theta \\
r^{2}  = x^{\prime 2}  +  \frac{y^{\prime 2}}{(1-\epsilon)^{2}} &
\end{cases}
\end{equation}
where $(x_0,y_0)$ are the coordinates of the X-ray emission peak, $\theta$ is the position angle, and $\epsilon$ is the ellipticity. The $\beta$-model may now be defined as follows:

\begin{equation}
\label{eq_beta}
S(r) = S_{0} \bigg(1+\frac{r^{2}}{r_{\rm c}^{2}}\bigg)^{-3\beta+1/2}  + b,
\end{equation}
where $S_{0}$ is the central brightness, $r_{\rm c}$ the core radius, $\beta$ the shape parameter, and $b$ a residual background emission, assumed
to be constant across the FOV.

\begin{figure}
\centering
\includegraphics[width=0.5\textwidth]{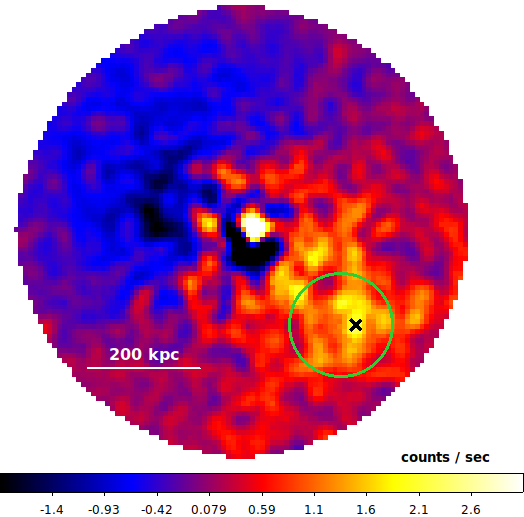}
\caption{Residual image after the 2D $\beta$-model subtraction. We marked the emission excess region and its peak inside the green circle.}
\label{unsharp_beta}
\end{figure}

 In the residual image, shown in Fig.~\ref{unsharp_beta}, there is an X-ray excess visible in the South-West direction, identified by the green circle. We mark the position of the peak of this residual emission excess by a black `x'. The projected distance between the X-ray peak of the cluster core to this emission excess peak is $\sim$ 1$'$. We have listed the best-fit parameters in Table \ref{beta_val}. In the residual image, the region of emission excess is larger and more scattered than in the previous Gaussian subtracted case. The residual emission is extending toward the North and South-East direction from its peak. This irregularity makes this region more asymmetric and complex around the emission excess peak. 

\subsection{Morphology parameters}
\par
\begin{center}
\begin{figure*}
    \centering
     \begin{subfigure}[t]{0.45\textwidth}
        \includegraphics[width=1\textwidth]{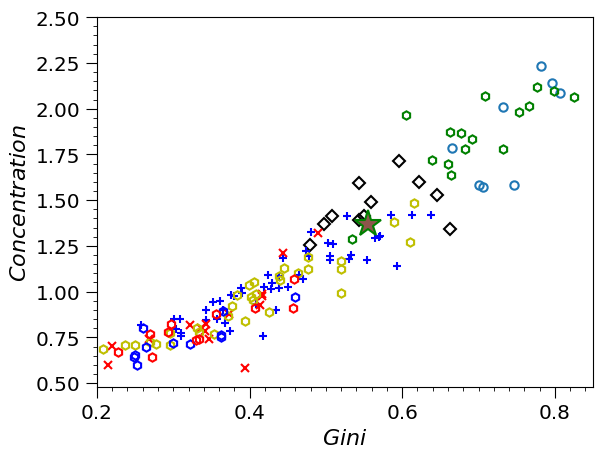}
        \caption{}
       \label{rfidtest_yaxis2}
     \end{subfigure}     
     \begin{subfigure}[t]{0.45\textwidth}
        \includegraphics[width=1.0\textwidth]{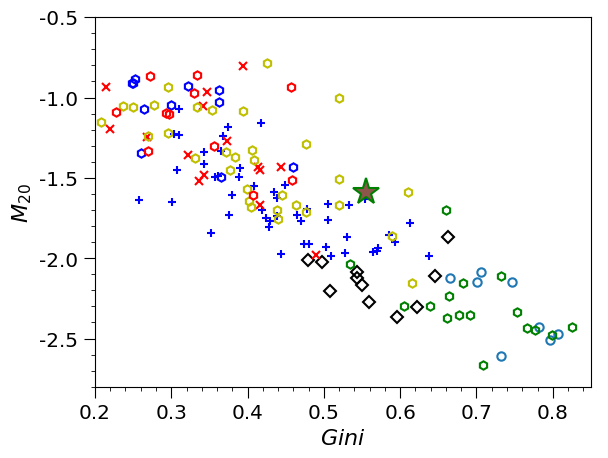}
        \caption{}
       \label{rfidtest_yaxis2}
     \end{subfigure}
     \begin{subfigure}[t]{0.45\textwidth}
        \includegraphics[width=1.0\textwidth]{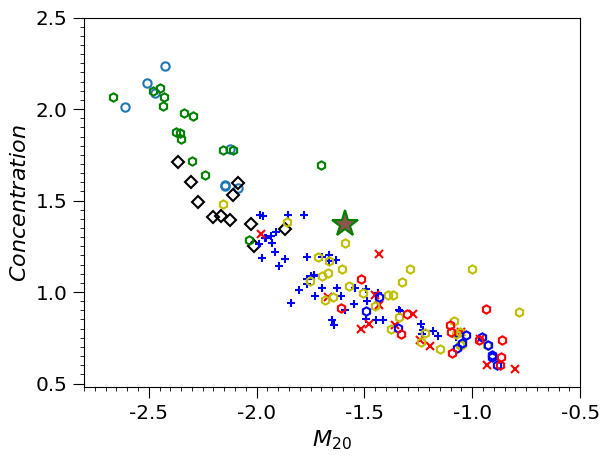}
        \caption{}
        \label{rfidtest_zaxis3}
     \end{subfigure}

     \caption[]{\small Three morphology parameters plotted in the parameter-parameter planes with
     RXCJ0232.2-4420 represented by a {\color{green} $\bigstar$}. $\circ$ = strong relaxed clusters, {\small $\lozenge$} = relaxed clusters, {\small \color {blue} $+$} = non-relaxed clusters, and {\small \color {red} $\times$} = strong non-relaxed. This sample is taken from \cite{2009ApJ...692.1033V} and morphology parameters calculated in \cite{2015A&A...575A.127P}. All clusters with exclusively diffuse radio sources are marked with hexagon. We took diffuse radio sources sample from \citep{2015ApJ...813...77Y}. Yellow hexagon correspond to only radio halo, blue hexagon to single relic, cyan hexagon correspond to double relics, and red hexagon to halo+relic and green hexagon to mini- halo.}
    \label{RXCJ_X-ray_morph_res}
\end{figure*}
\end{center}

 In this section, we analyse the X-ray morphology of RXCJ0232.2-4420 to characterise the degree of disturbances in the projected X-ray emission. This analysis will also allow us to investigate the dynamical state of the cluster and its comparisons with other X-ray clusters' morphologies. For this work, we computed three morphology parameters - Gini, $M_{20}$ and concentration - using RXCJ0232.2-4420 {\it Chandra} X-ray image in the [0.5-7 keV] energy band. These morphology parameters trace the flux distribution for a given region, and their computed values depend on the degree of disturbance in a given cluster image. For more details about the parameters and method, we refer to the \cite{2015A&A...575A.127P}. We calculate these parameters for a circular region of 500 $h_{70}$ kpc of radius, centred on the flux weighted centroid (see Table \ref{RXCJ_X-ray_morph}).  Fig.~(\ref{RXCJ_X-ray_morph_res})
shows these three morphology parameters in the parameter-parameter planes. We represent RXCJ0232.2-4420 along with a sample of relaxed and merging clusters taken from \cite{2009ApJ...692.1033V},
for which their morphology parameters were computed in our previous work \citep{2015A&A...575A.127P}. In the same plot, we have also shown a sample of diffuse radio clusters, including radio halos, relics and mini-halos \citep{2015ApJ...813...77Y}.

\begin{table}
\centering
\caption{RXCJ0232.2-4420 morphology parameter values.}
\begin{center}
\begin{tabular}{@{}cccccccc@{}}
\hline
\hline 
&Gini &$M_{20}$ & concentration \\
\hline

& 0.55 $\pm$0.005 & -1.58 $\pm$ 0.046 & 1.37 $\pm$ 0.014\\

\hline 
\end{tabular}
\end{center}
\label{RXCJ_X-ray_morph} 
\end{table}

\subsection{Temperature distribution}
\par In order to understand the temperature distribution, we measure the temperature for different regions of the RXCJ0232.2-4420 cluster. First of all, we study the azimuthal mean temperature measurements. \cite{2009ApJS..182...12C} have derived the radial temperature of RXCJ0232.2-4420 using the {\it Chandra} data\footnote{https://web.pa.msu.edu/astro/MC2/accept/clusters/4993.html}. They used five radial bins (each contains 2500 counts) up to 500 kpc radius centred on the X-ray peak.

\par
\begin{figure}
\centering
\includegraphics[width=0.5\textwidth]{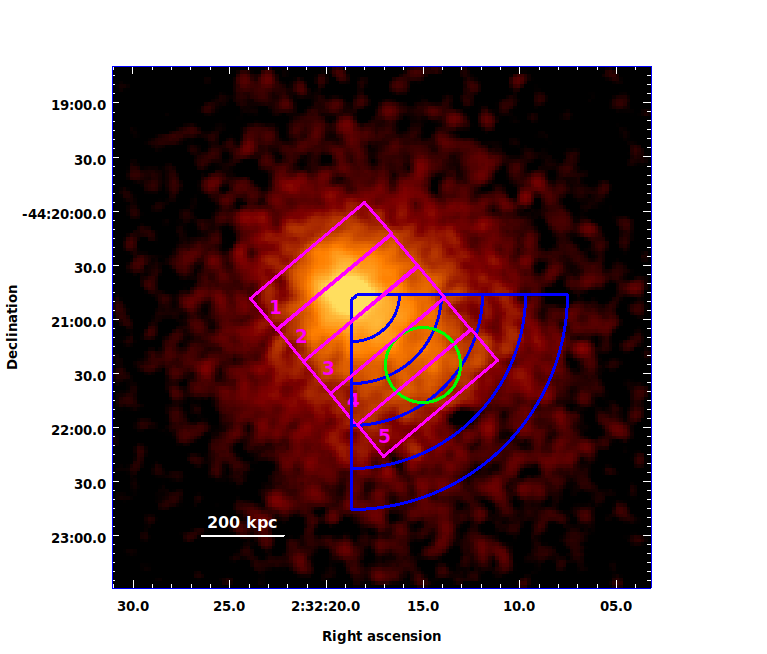}
\caption{Spectra extracted from the various regions of RXCJ0232.2-4420.}
\label{temp_reg}
\end{figure}

\begin{center}
\begin{figure*}
    \centering
    \begin{subfigure}[t]{0.45\textwidth}
        \includegraphics[width=1\textwidth]{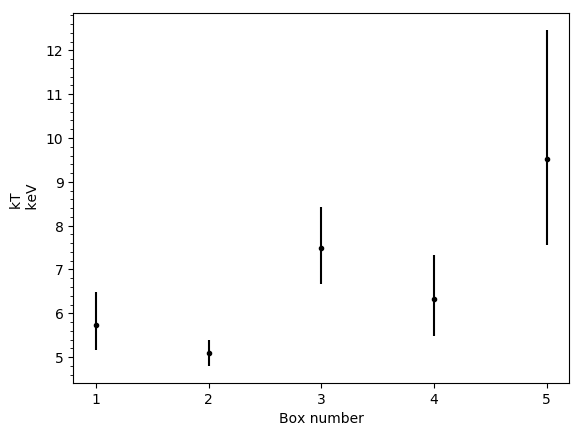}
        \caption{}
        \label{rfidtest_xaxis1}
    \end{subfigure}
    \begin{subfigure}[t]{0.45\textwidth}
        \includegraphics[width=1\textwidth]{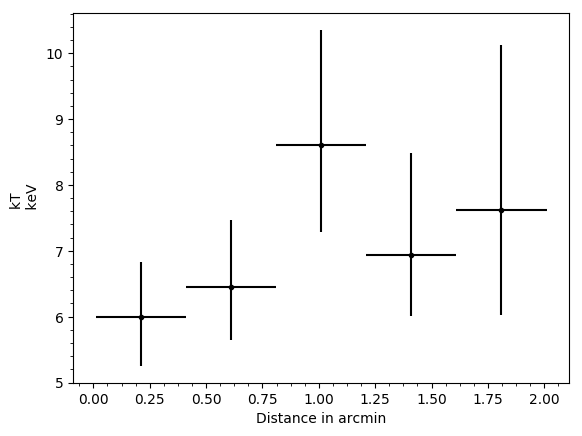}
        \caption{}
        \label{rfidtest_yaxis2}
        \end{subfigure}
    \caption{(a) Temperature plotted for the box regions. (b) Temperature plotted within the sector 4 bins vs the distance from the core.}
    \label{temp_prof_reg}        
\end{figure*}
\end{center}

 Apart from the azimuthal radial profile, we also study the temperature distribution; (1) across the North-East to South-West cluster emission by dividing the cluster into five boxes and (2) within the sector 4 (Fig.~\ref{SB_prof_sec}(a)) which is the 270$^{\circ}$ to 360$^{\circ}$ angle. We show both of these regions in Fig.~\ref{temp_reg}. In addition to this, we also extract the spectra from the emission excess region (Fig.~\ref{unsharp} green circle). We plot the temperature profiles for both of these regions (boxes and sector) in Fig.~\ref{temp_prof_reg}. In \$\ref{sb}, we notice that the surface brightness profile in sector 4 is deviating from the other profiles. Hence, we extract the spectra from sector 4 up to 2$'$ maximum radius from the cluster core within five bins. This is shown in Fig.~\ref{temp_reg} and corresponds to the temperature profile in Fig.~\ref{temp_prof_reg}(b).  



\begin{table}
\caption{Best-fitting spectral values of the of the box regions as shown in Fig.~\ref{temp_reg}.}
\begin{tabular}{ccccccccc}
\hline
\hline
Region & T &Normalisation & $\chi^{2}$/dof\\
& keV &  $\times$10$^{-3}$ $cm^{-3}$ & &  \\
\hline
1& 5.74$^{+0.74}_{-0.56}$ &  0.840$^{+0.031}_{-0.031}$ & 80/118    \\
2& 5.09$^{+0.30}_{-0.29}$ &  2.284$^{+0.052}_{-0.052}$ & 164/204   \\
3& 7.48$^{+0.95}_{-0.82}$ &  1.086$^{+0.033}_{-0.033}$ &  93/152      \\
4& 6.32$^{+1.01}_{-0.85}$ &  0.747$^{+0.032}_{-0.031}$ &  66/101      \\
5& 9.52$^{+2.95}_{-1.96}$ &  0.442$^{+0.021}_{-0.021}$ &  40/74       \\

\hline
\end{tabular}
\label{spectral_val}
\end{table}

\begin{table}
\caption{Best-fitting spectral values of the pie regions (sector 4) as shown in Fig.~\ref{temp_reg}.}
\begin{tabular}{ccccccccc}
\hline
\hline
Bins & T &Normalisation & $\chi^{2}$/dof\\
& keV &  $\times$10$^{-3}$ $cm^{-3}$ & &  \\
\hline
1&6.00$^{+0.83}_{-0.75}$ &  0.736$^{+0.029}_{-0.030}$ &   63/110\\
2&6.46$^{+1.01}_{-0.81}$ &  0.771$^{+0.031}_{-0.031}$ &   65/113\\
3&8.61$^{+1.75}_{-1.32}$ &  0.691$^{+0.028}_{-0.026}$ &   60/104\\
4&6.94$^{+1.54}_{-0.93}$ &  0.469$^{+0.022}_{-0.022}$ &   54/79\\ 
5&7.63$^{+2.49}_{-1.61}$ &  0.345$^{+0.021}_{-0.021}$ &   29/57\\ 

\hline
\end{tabular}
\label{spectral_val_2}
\end{table}

\subsection{Thermodynamic maps}

\begin{center}
\begin{figure*}
    \centering
    \begin{subfigure}[t]{0.45\textwidth}
        \includegraphics[width=1\textwidth]{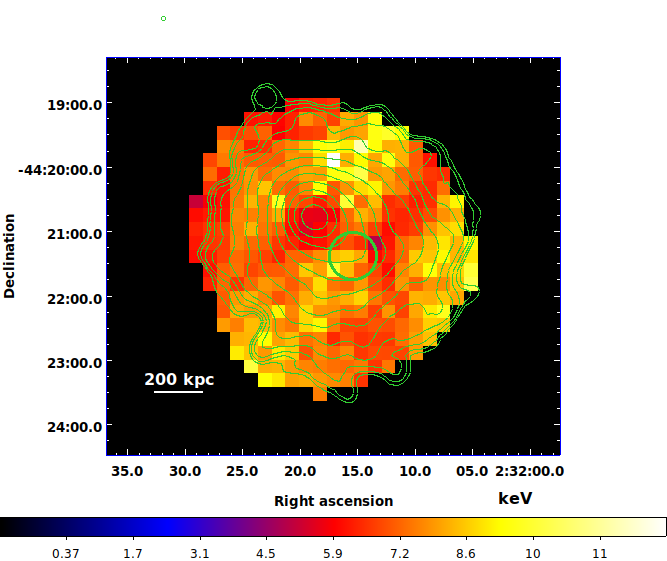}
        \caption{}
        \label{rfidtest_xaxis1}
    \end{subfigure}
    \begin{subfigure}[t]{0.45\textwidth}
        \includegraphics[width=1\textwidth]{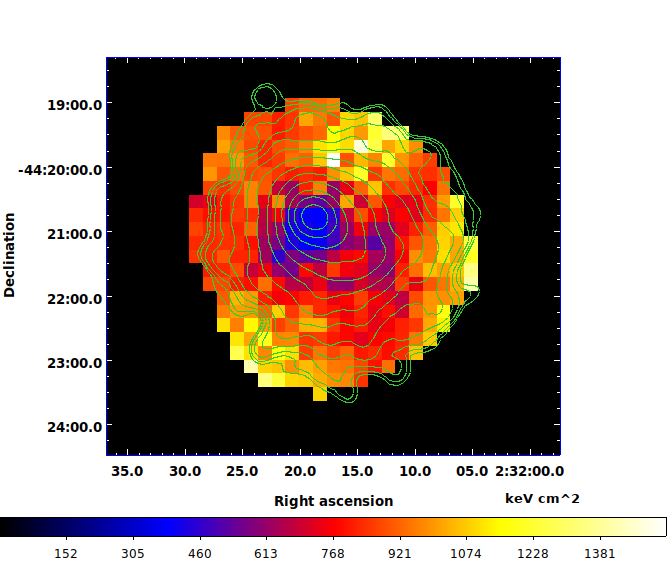}
        \caption{}
        \label{rfidtest_yaxis2}
        \end{subfigure}
    \begin{subfigure}[t]{0.45\textwidth}
        \includegraphics[width=1\textwidth]{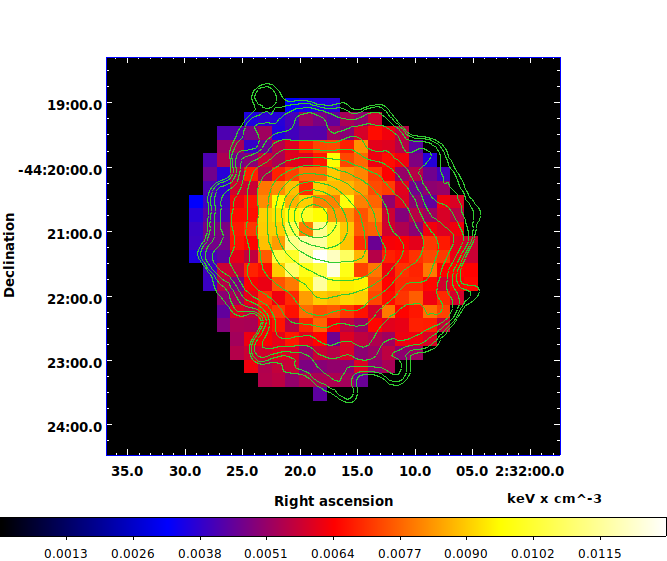}
        \caption{}
        \label{rfidtest_xaxis1}
    \end{subfigure}
    
    \caption{Thermodynamic maps of RXCJ0232.2-4420. (a) Temperature ($kT$) map. (b) Pseudo-entropy ($S$) map. (c) Pseudo-pressure ($P$) map. In all maps, {\it Chandra} smoothed image contours are overlaid. }
    \label{thermo_map}        
\end{figure*}
\end{center}

\par in Fig.~\ref{thermo_map}, we show the 2D temperature ($kT$), pseudo-entropy ($S$) and pseudo-pressure ($P$) maps for the cluster RXCJ0232.2-4420. {We show temperature error and reduced $\chi^{2}$ maps into Appendix Fig. 1}. This was achieved by dividing the data into small regions from which spectra were extracted \citep[this procedure was already described in][]{Durret10, Durret11, Lagana08, Lagana19}. {We set a minimum count number of 900 net counts (that is, after background subtraction)} to each pixel and, if necessary, we increased the region up to a box of 5 $\times$ 5 pixels. Each pixel has a size of 12.8 $\times$ 12.8 arcsec$^2$, corresponding to 54.8 $\times$ 54.8 $h_{70}$ kpc$^2$. 
If we still do not have sufficient counts after binning, the pixel is ignored, and we proceed to the next neighbouring pixel. When we have adequate counts, the spectra of  MOS1, MOS2 and pn are then simultaneously fitted, and the best-fit $kT$ and $Z$ values are attributed to the central pixel. 

\section{Discussion}

\subsection{Radio halo, morphology and substructure}
\par {RXCJ0232.2-4420 is a peculiar system hosting a radio halo of the size 550 $\times$ 800 kpc$^{2}$ \citep{2019MNRAS.486L..80K}. The size of the halo is intermediate between large radio halos and mini halos.
In line with the characteristic of mini-halos to occur around BCGs, and unlike typical radio halos in merging clusters where the BCG is often displaced from the centre, this radio halo surrounds the BCG in the cluster. }
From a visual inspection, its X-ray image or surface brightness distribution does not show an obvious departure from the thermodynamic equilibrium, unlike other merging systems. {Thus, it was classified as a system in transition from mini-halos to radio halos.} The cluster is elongated in the North-East to the South-West direction, similar to the A1995 merging system \citep{2012A&A...547A..44B} in which a diffuse radio halo {has been} detected. For the A1995 merging cluster, the authors used X-ray and optical data and identified two main optical substructures causing the NE-SW elongation of the galaxy distribution what suggests a bimodal merger caught in a phase of post core-core passage.

\par RXCJ0232.2-4420 cluster was part of the Archive of Chandra Cluster Entropy Profile Tables (ACCEPT) sample \citep{2009ApJS..182...12C}. They estimated the entropy and cooling time of the central (30 kpc) region to be $\sim$ 50 keV cm$^{2}$ and 1 Gyr, respectively. \cite{2008A&A...483...35S} classified as strong cool-core clusters those having cooling time $<$ 3 Gyr, which puts RXCJ0232.2-4420 in this category. Further, \cite{2010A&A...513A..37H} have measured the central entropy for HIFLUGCS (HIghest X-ray FLUx Galaxy Cluster Sample) cluster sample and subdivide clusters into (i) strong cool-core, (ii) weak cool-core or (iii) no cool-core clusters. Based on their analysis, they found the strong cool-core clusters have central entropy $<$ 30 keV cm$^{2}$, and no cool-core clusters have central entropy $>$ 110 keV cm$^{2}$. According to this classification, the RXCJ0232.2-4420 cluster falls into the weak cool-core cluster. 

\par We study the azimuthal average surface brightness profile over 4$'$ region centred on the peak of the X-ray emission. We could see the departure {from} the $\beta$-model at $\sim$ 1$'$ from the fitting (Fig.~\ref{SB_prof}), and it suggests disturbances in surface brightness distribution. To investigate this disturbance, we divided the cluster emission into four sectors and studied the profiles in each of the sectors (Fig.~\ref{SB_prof_sec}). Out of the four sectors, there is one sector that clearly shows the disturbances in the South-West direction, which is the elongation axis of the cluster. This sector spans the 270-360$^{\circ}$  area and hosts the possible substructure. To further investigate this substructure, we use two different filtering techniques. One is simple unsharp masking which can be used to amplify the high-frequency components of the image. To do so, we adopt four different Gaussian kernels (as shown in Fig.~\ref{unsharp}), and in all of the residual images, we detect an X-ray emission excess in the South-West direction. We consider this to be an underlying substructure associated with RXCJ0232.2-4420. The distance from the peak (or BCG) to this substructure is $\sim$ 1$'$. We measure the size of this substructure to be 0.35$'$ or 90 $h_{70}$ kpc of the radius. Similarly, we also modelled the cluster emission with the 2D $\beta$-model and subtracted it from the original image. This residual image (Fig.~\ref{unsharp_beta}) clearly depicts the substructure, much extended than what is seen in the unsharp masked image. Both techniques reveal the presence of substructure in the South-West direction where the cluster is elongated.

\par To add to our analysis, we study the three morphology parameters, namely, Gini, $M_{20}$ and concentration, to compare the morphology of RXCJ0232.2-4420 with other samples of clusters with different dynamical states.  Following the classification criteria provided by \citet{2015A&A...575A.127P} for the dynamical state based on these three parameters, the  strong relaxed (strong cool-core) clusters have Gini $>$ 0.65, concentration $>$ 1.55 and $M_{20}$ $<$ -2.0, while the disturbed clusters have Gini $<$ 0.40, concentration $<$ 1.0 and $M_{20}$ $>$ -1.4. All other clusters falling in the intermediate dynamical states (or weak cool core) have the parameters between these ranges. {The spatial parameter values obtained for RXCJ0232.2-4420 do not place this cluster clearly in any particular category, besides displaying some level of the substructure.} We showed the position of RXCJ0232.2-4420 along with other samples of clusters in Fig.~\ref{RXCJ_X-ray_morph_res}.
 In this figure, the position of RXCJ0232.2-4420 also matches the single radio halo clusters. Overall X-ray morphology indicates that the RXCJ0232.2-4420 is a slightly disturbed cluster that is elongated in one particular direction and has a bright core. The flux weighted centroid does not coincide with the X-ray peak, falling on the elongated South-West axis below the X-ray peak (17$''$ away). 
\par The low cooling time of the core of RXCJ0232.2-4420 and spatial coincidence between X-ray peak and BCG may suggest that the core of the cluster might not be disturbed by a cluster merger yet. \cite {2007A&A...463..839R} analysed the {\it Chandra} and {\it XMM} observation of A3558 and found that the cool-core of the A3558 had survived in a merger. \cite {2006ApJ...643..751C} showed that the A2065 is an ongoing merging cluster that might be a possible reason behind the survival of one cool-core in this unequal mass merger. Several observations and simulations show that the elongation of a cluster is the result of the accretion of groups along filaments \citep{1998A&A...335...41D,1997ApJS..109..307R}. RXCJ0232.2-4420 is showing that the possible interaction with other groups or clusters has been in the elongated direction, but the merger shocks have not reached the core of the cluster. Hence, the core presents low temperature (and entropy) and cooling time values as its cool-core has still survived in the merging process. 

\subsection{Temperature distributions}
\par {RXCJ0232.2-4420 is a hot and luminous cluster. \citet{2009ApJS..182...12C} estimated the temperature over a 5.94$'$ region (which corresponds to a linear size of $\sim$ 1.5 Mpc) {to be $kT$ $\sim$ 7.8 keV} and its corresponding bolometric luminosity is $L_{\rm X} = 18.30 \times 10^{44}$ erg/s}. Generally, clusters with radio halos are hotter and luminous \citep[e.g.][]{2012A&ARv..20...54F}. It was observed that about 30\% of the high X-ray luminosity ($L_{\rm X}$ $>$ 5$\times$10$^{44}$ erg s$^{-1}$) galaxy clusters host radio halos. Furthermore, we note that the simulations and observations of merging galaxy clusters have shown that mergers can temporarily boost the X-ray temperature and luminosity of the systems \citep{2004JKAS...37..433S,Randall_2002}. 
{The underlying substructure could cause the high temperature of RXCJ0232.2-4420}.
{Its radial temperature profile \citep{2009ApJS..182...12C} suggests a lower temperature of $\sim$ 5-6 keV in the inner core ($<$ 50 kpc) region, rising to $\sim$ 8-9 keV at the distance of 200 kpc from the centre where the identified substructure is located}. Going towards the outskirts, the derived temperature has large error bars and thus becomes unreliable. We also study the temperature distribution across the North-East to South-West direction, dividing it into five boxes of size 354 $\times$ 97 $h_{70}$ kpc$^{2}$, as shown in Fig.~\ref{temp_reg}. As can be seen in the temperature profile of the box regions (Fig.~\ref{temp_prof_reg}(a)), the temperature is low, $\sim$ 5-5.7 kev, in boxes 1 and 2, corresponding to the inner regions of RXCJ0232.2-4420, close to the BCGs. Then temperature rises up from $\sim$ 6.3 to 9.5 keV into the boxes numbered 4 and 5. {It is important to state that the errors associated with temperature measurements are large (but temperatures from region 4 and 5 still disagree within 1-$\sigma$), hence the statistical significance of results should be evaluated carefully.} These boxes are spatially coincident with the position of the substructure. We also study the temperature profile in the sector between 270-360$^{\circ}$ angle centred on the X-ray peak. Similarly to the box profile, the temperature is low in the inner bins close to the BCGs. At the distance of $\sim$ 1$'$ temperature rises to $\sim$ 8.5 keV, and then it decreases again. This suggests the substructure is surrounded by low-temperature gas.  We measured the temperature of the substructure region, indicated by the green circle in Fig.~\ref{temp_reg}, {and it is $kT = 8.82 \pm 2.16$ keV.} This analysis suggests that the presence of this substructure could affect the global X-ray properties of the cluster. 

\subsection{Thermodynamic properties}
\par We also perform the analysis of 2D thermodynamic maps of RXCJ0232.2-4420 using XMM-\textit{Newton} data. As shown in Fig.~\ref{thermo_map}, the pseudo-pressure map shows the high-pressure fluctuations in the South-West direction near the position of the substructure as detected in the {\it Chandra} residual map. We marked the position of the substructure in Fig.~\ref{thermo_map}(a) with the green circle. We note that in this (XMM-\textit{Newton}) pressure map, the position of the substructure as revealed by the {\it Chandra} image does not coincide with the South-West high-pressure region. We found there is an offset of $\sim$ 30$''$ between the substructure regions in {\it Chandra} and XMM-\textit{Newton}. 
{The possible reason behind the disagreement between substructure position could be the lower spatial resolution of XMM-\textit{Newton} telescope. 
XMM-\textit{Newton} temperature map shows the high temperature ($\sim$ 8-9 keV) fluctuations (at the location of high pressure) which is surrounded by the low-temperature gas of $\sim$ 6-7 keV. This is supported by the temperature profile in sector 4 (Fig.~\ref{temp_prof_reg}(b)), where the temperature is dropping outside the substructure region. The pseudo-entropy map also shows the fluctuations at the location of the high-pressure region. There is a low entropy gas near the cluster core, but high entropy gas extends towards the substructure position. The fluctuation in the temperature and entropy maps could be the result of a mixing of hot gas between RXCJ0232.2-4420 and its substructure.}


\section{Conclusion}
\par {RXCJ0232.2-4420 is a hot (of $kT$ $\sim$ 7.8 keV) and luminous ($L_{\rm X} = 18.30 \times 10^{44}$ erg/s) cluster with a small scale (of 180$h_{70}$ kpc diameter) substructure}. One of the characteristics of the system is the transition phase of its radio halo from mini-halo to giant radio halo \citep{2019MNRAS.486L..80K}. In a typical mini-halo system, the core of the host cluster shows disturbances in the form of cold fronts, swirling of the hot gas or AGN activity \citep{2015aska.confE..76G,2011MmSAI..82..632Z}. These disturbances may provide turbulent re-acceleration to the seed relativistic particles. While in the RXCJ0232.2-4420 case, there are no disturbances detected at the core in X-ray data,
there is a faint substructure located $\sim$ 1$'$ away from the core that 
could have survived in a past merging process. Furthermore, this substructure could be in the process of injecting turbulence in the RXCJ0232.2-4420 system and hence slowly power the radio halo due to the particle re-acceleration. This indicates the radio halo is still growing around the core of the RXCJ0232.2-4420 cluster. Our future MeerKAT and uGMRT wideband and more sensitive radio observations will shed light on the nature of the radio halo of RXCJ0232.2-4420, as well as how substructure affects the magnetic field distributions.



\section*{Acknowledgements}
The financial assistance of the South African Radio Astronomy Observatory (SARAO) towards this research is hereby acknowledged (www.ska.ac.za). This work is based upon research supported by the South African Research Chairs Initiative of the Department of Science and Technology and National Research Foundation. TFL acknowledges financial support from the Brazilian agencies FAPESP and CNPq  through grants 2018/02626-8 and 306163/2019-5, respectively. RK acknowledges the support of the Department of Atomic Energy, Government of
India, under project no. 12-R\&D-TFR-5.02-0700. We thank the staff of the GMRT who have made these observations possible. The GMRT is run by the National Centre for Radio Astrophysics of the Tata Institute of Fundamental Research.

\section*{Data Availability}
The data underlying this paper are available in {\it Chandra} and XMM-{\it Newton} telescopes archive systems, at https://cda.harvard.edu/chaser/ and https://nxsa.esac.esa.int/nxsa-web/.

\bibliography{references}

\appendix
\section*{Appendix}

\begin{figure}[thb]
    \centering
    \begin{subfigure}{0.45\textwidth}
        \includegraphics[width=1\textwidth]{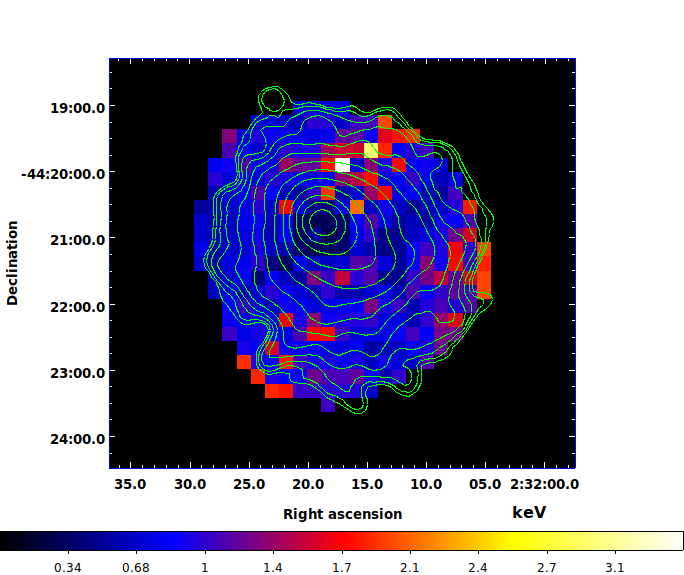}
        \caption{}
        \label{rfidtest_xaxis1}
    \end{subfigure}
    \begin{subfigure}{0.45\textwidth}
        \includegraphics[width=1\textwidth]{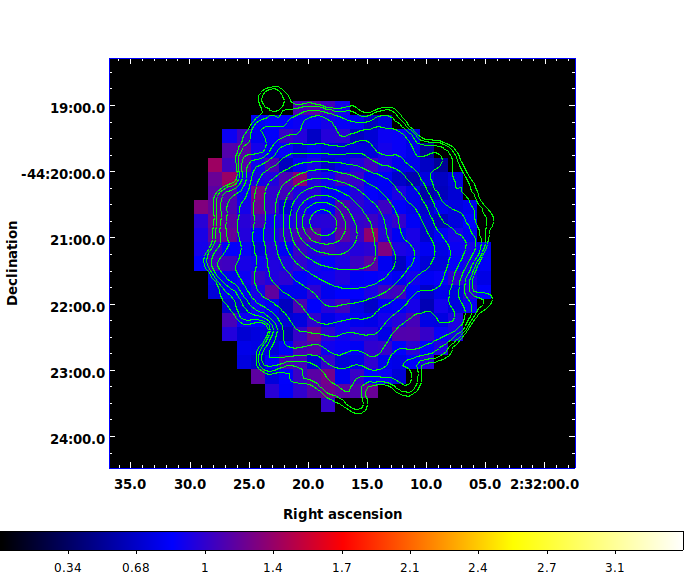}
        \caption{}
        \label{rfidtest_yaxis2}
        \end{subfigure}
    \caption{(a) $kT$ error and (b) { reduced} $\chi^{2}$ maps corresponding to XMM-\textit{Newton} 2D $kT$ map. In all maps, {\it Chandra} smoothed image contours are overlaid. }

\end{figure}

\end{document}